# Test for consistence of a flyby anomaly simulation with the observed Doppler residuals for the Messenger flybys of Mercury

H. J. Busack

Wulfsdorfer Weg 89, 23560 Lübeck, Germany
07. March 2010

Abstract

In 2007, the observed Earth flyby anomalies have been successfully simulated using an empirical formula (H. J. Busack, 2007). This simulation has led to the prediction of anomaly values, to be expected for the velocity changes during Rosetta flybys of Mars in 2007, and following twice of Earth in 2007 and 2009. While the data for the Mars flyby are yet under evaluation, the predictions of the used formula for the last two Earth flybys of Rosetta are fully confirmed now. This is remarkable, since an alternatively proposed formula (Anderson et al., 2007) failed to predict the correct values for the recent flybys. For the Mercury flybys of the Messenger spacecraft, this alternative formula predicts a null result. In the meantime, Doppler residuals of these flybys on 14.01.2008 and 06.10.2008 are available. On both flybys significant residuals were observed, using gravity data derived by Mariner 10 on Mercury (D.E. Smith et al., 2009). According to the authors, these residuals cannot be eliminated totally by adjustment of the second degree gravity coefficients and by assumption of irregular mass concentrations of acceptable value on Mercury. In this investigation, I adapt the output of the simulation program to compare with the measured Doppler residuals of the Mercury flybys without changing the formerly derived parameters for the Earth flybys. The simulation with these parameters leads to Doppler residuals of the Mercury flybys compatible with the measured curves. Additionally, the expected flyby anomalies are calculated. Since the gravity field of Mercury is not explored yet with sufficient accuracy, this result cannot be falsified or confirmed until the evaluation of the coming Mercury orbits of Messenger will be finished. If the proposed empirical formula would be confirmed then again, this would be a strong indication of an underlying physical reality.

## 1   Introduction

Since the exploration of anomalous velocity changes during the Earth flybys of several spacecrafts, many efforts have been made to explain these anomalies by unconsidered effects of known physics [5, 6, 7, 8] or by changing some properties of known physics [11, 12, 13, 14, 15]. None of these efforts were able to explain the effects completely.

Additionally, two empirical formulas have been proposed [1, 3], to explain the anomaly at least phenomenologically. While both formula described the previously measured anomaly values on an acceptable accuracy level, they have led to different predictions for the anomaly values at infinity of the second and the third Rosetta flyby of Earth on 13.11.2007 and 13.11.2009, respectively.

The formula given by Busack [1] predicts the following anomaly values for the Rosetta flybys:

Rosetta 2:  ( 0.00 +/- 0.01) mm/s
Rosetta 3:  ( 0.00 +/- 0.04) mm/s

The formula by Anderson et al. [3] predicts (without error limit) the following anomaly values:

Rosetta 2:   0.98 mm/s
Rosetta 3:   1.09 mm/s

The measured values at perigee are [2]:

Rosetta 2:   0 mm/s (no error limit provided)
Rosetta 3:   (-0.004 +/- 0.044) mm/s

This can be recalculated for values at infinity:

Rosetta 2:   0 mm/s
Rosetta 3:   (-0.006 +/- 0.06) mm/s

The measured data clearly show perfect congruence with that predicted by Busack. While the null result of the second flyby may have been attributed to the perigee altitude of 5322km rather than been considered as evidence of the simulation formula, the third flyby had a perigee altitude of only 2483km, comparable with the altitude of the first flyby (1954km). On this first flyby a significant velocity increase of 1.82mm/s has been seen, the null result of the second flyby therefore has to be considered as confirmation of the simulation formula.
This fact was encouraging for further investigations on the recent Messenger flybys of Mercury, presented here.

D.E. Smith et al. recently have presented a paper [4], showing the Doppler residuals estimated for the first Mercury flybys on 14.01.2008 and 06.10.2008. For the third flyby on 29.09.2009, Doppler data in the vicinity of perigee was not obtainable because of lost of solar power in this mission phase.
Using the gravity data of mercury as derived by Mariner 10, Smith et al. calculated residual data which show significant perturbations of several cm/s in the vicinity of closest approach in both cases. In the case of the first flyby, the foregoing data are not obtainable for a period of about 47 minutes because of occultation of the spacecraft by Mercury in Earth direction.
The perturbations could be minimized but not fully eliminated by adjustment of the second degree gravity data, therefore beeing different from the Mariner 10 data, and by assumption of irregular mass concentrations along the ground track and within certain impact basins. Smith et al. stated that the residual patterns in this case could be reduced to 17mm/s and 16mm/s, respectively. Unfortunately, they gave no figure of the resulting residual curves.
Equally, there are no statements regarding the resulting velocity change at infinity, normally taken as measure for the flyby anomaly. Of course, this could be done accurately only with better gravity data. These data are hopefully obtained by analyzing the Mercury orbits of Messenger beginning in 2011.

According to the Anderson formula, no measurable anomaly should be explored in both flybys, mainly due to the low angular momentum of Mercury. But even in the case, one considers the same factor of the cosine term of the formula as for the Earth, the predicted anomaly would be less than 1mm/s in both cases (-0.4mm/s and -0.7mm/s, respectively) .

The Busack formula leads with the parameter sets "low" and "high" finally derived for the Earth flybys to the following anomaly values at infinity for both Mercury flybys:

14.01.2008:  -9.9mm/s "low" ,  -6.4mm/s "high"
06.10.2008:  -7.3mm/s "low" ,  -4.9mm/s "high"

Since flyby anomaly values at infinity are not provided by Smith et al., the computer program used by Busack for the calculations, was modified in such a way, that Doppler residuals equivalent to those calculated by Smith et al. could be derived.

## 2  Background of the Doppler residual simulation

Justification for the used formula was given in [1]. For reference purposes this formula is reproduced here in the finally used form.

$$\vec{g}(\vec{r}) = -\frac{G \cdot M \cdot \vec{r}}{r^3} \left[ 1 + A \cdot \exp\left( -\frac{r - R}{B - C \dfrac{\vec{r} \cdot \vec{v}}{r \cdot v_{Sun}}} \right) \right] \quad (1)$$

with $r = |\vec{r}| \geq R$ and

- $G$: gravitation constant
- $M$: field mass
- $R$: radius of the field mass body
- $\vec{r}$: position vector of the test mass with origin at the center of the field mass body
- $\vec{v}$: velocity vector of the field mass center in the gravitational rest frame
- $v_{Sun}$: magnitude of the Sun velocity in the gravitational rest frame
- $A,B,C$: arbitrary constants

For clarity purposes this could be transformed to

$$\vec{g}(\vec{r}) = -\frac{G \cdot M \cdot \vec{r}}{r^3} \left[ 1 + A \cdot \exp\left( -\frac{h}{B\left(1 - C' \dfrac{\vec{r} \cdot \vec{v}}{r \cdot v_{Sun}}\right)} \right) \right] \quad (1a)$$

with h = altitude and C'=C/B.

The free parameter A is the amount of the anomalous acceleration at h=0, the parameter B defines the slope of the decrease with altitude, and C'=C/B defines the amount of asymmetry in direction of motion within the assumed gravitational rest frame. The asymmetry introduced by this term is approximately of harmonic first order in direction of the apex of motion in the rest frame.

Interestingly, this implies, that gravity determinations in most cases would depend on altitude more than expected by known physics, and could show a diurnal and annual variance for fixed observation sites on the body surface. Those effects possibly have been observed [9].

The parameters used for the final evaluation of the Earth flybys of all investigated spacecrafts and for the here analyzed Mercury flybys as well, were:

|              | Low       | high      |           |
|--------------|-----------|-----------|-----------|
| A            | : 0.0002453 | 0.0002306 |           |
| B / m        | : 394000  | 464000    |           |
| C'           | : 0.3452  | 0.2263    |           |
| Vsun / m/s   | : 360000  | 360000    |           |
| RA / h       | : 17.78   | 17.7      | \| apex   |
| DEC / degree | : -37.5   | -61       | \| direction |

Obviously, due to the dot product in the formula, the resulting accelerations differ in magnitude and direction from those derived from Newton's law. Therefore, the resulting asymptotic track can differ in any direction from the Newtonian track, dependent on the track velocity with regards to the apex direction in the vicinity of closest approach. Equally, even in the case of a perfectly known gravity field, those differences in range and Doppler signal to expected tracks can occur, if the impact parameter is not known to sufficient accuracy. For this reason, the impact parameter, or more accurately, the center coordinates of the mass body have to be varied until the range and Doppler data of the calculated asymptotic spacecraft track match the measured data. Residuals observed with this procedure are an indication, that the gravity field of the mass body is not modelled yet with sufficient accuracy, based on known physics, or that additional effects have to be taken into account. These effects could be of non gravitational nature, not covered by the orbit determination program, or gravitational effects not covered by known physics.

In this investigation, the latter way was analyzed by using the empirical gravity term introduced by Busack. The program previously used for simulation of the flyby anomaly [16] is something like a rudimental orbit determination program, added with equation (1). The free parameters have been varied in [1] in order to match the measured flyby anomaly values. For the purpose of this paper, the program was modified by adding the data of the Messenger Mercury flybys, by adding the position vector from Earth to Mercury for the times of both flybys, and by means for adjusting the center coordinates and the mass of the mass body for the Newtonian track . The position vector from Earth was used to determine the Doppler velocity components and the range components of the calculated tracks, since they are measured from Doppler ranging stations on Earth. The parameters previously derived for simulating the Earth flybys were left unchanged.

## 3   Realization of the Doppler residual simulation

The simulation has been written as a computer program by means of a high-level programming system of the language family PASCAL. Details of the basic function can be found in [1].

The track data of the Messenger flybys of Mercury and the position vectors from Earth to Mercury for both flybys were taken from the HORIZONS web interface of JPL [10]. The difference data for Doppler velocity and range were derived from the calculated velocity and position vectors by scalar multiplication with the Earth direction unity vector. By variation of the center coordinates of Mercury for the Newtonian track, the asymptotic Doppler and range residuals to the non Newtonian track were minimized. The variation of the center coordinates was of the order of one hundred meters. This procedure is somewhat arbitrary, because there are different ways to do this. The way with the smallest amplitude for the residuals at closest approach was chosen. This has been regarded to be an adequate comparison to the Doppler residuals derived by Smith et al..

For reference purposes, a download of the executable program file flyby_anomaly2.exe is provided by [17]. The source code can be requested from the author.

## 4   Results of the simulation

Fig.1 and Fig.2 show the comparison of the simulated Doppler residual curves with those derived by Smith et al. for a Mercury gravity field as known from the Mariner 10 flybys in the $70^{th}$. The simulations shown are derived with parameter set "low". Parameter set "high" yields quite similar curves. Of course the simulated curves can not match the measured curves completely, because the Mercury gravity field is not known to sufficient accuracy for such comparison. As Smith et al. stated, for plausible changes in the second degree gravity coefficients and with plausible distribution of local gravity anomalies the doppler residuals could be minimized. Especially the striking residuals of the second flyby could be drastically reduced. They could decrease the perturbations to about 17mm/s for the first flyby and to 16mm/s for the second flyby. This in mind, one can consider the congruence of the curves as noticable.

**Fig.1** : Measured and simulated Doppler residuals of the first Messenger flyby of Mercury. Measured values after [4] for a gravity field of Mercury from Mariner 10 data.

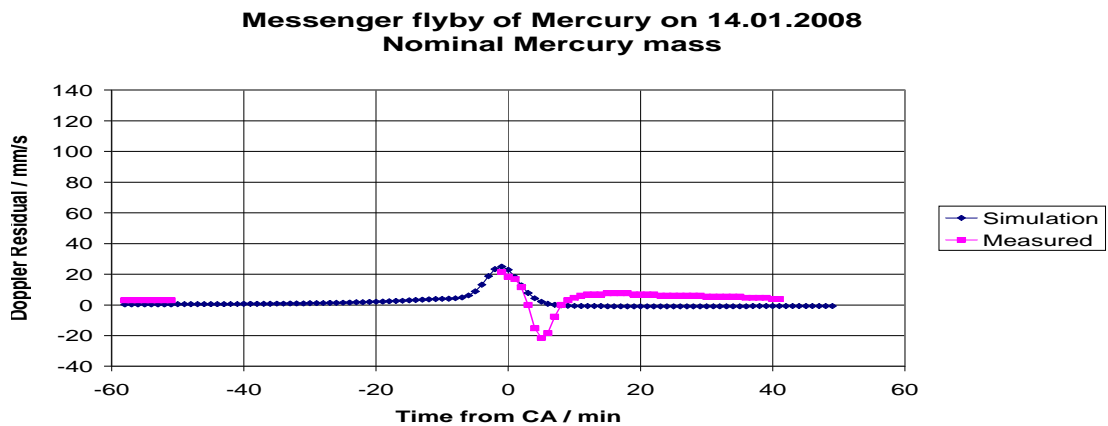

**Fig.2** : Measured and simulated Doppler residuals of the second Messenger flyby of Mercury. Measured values after [4] for a gravity field of Mercury from Mariner 10 data.

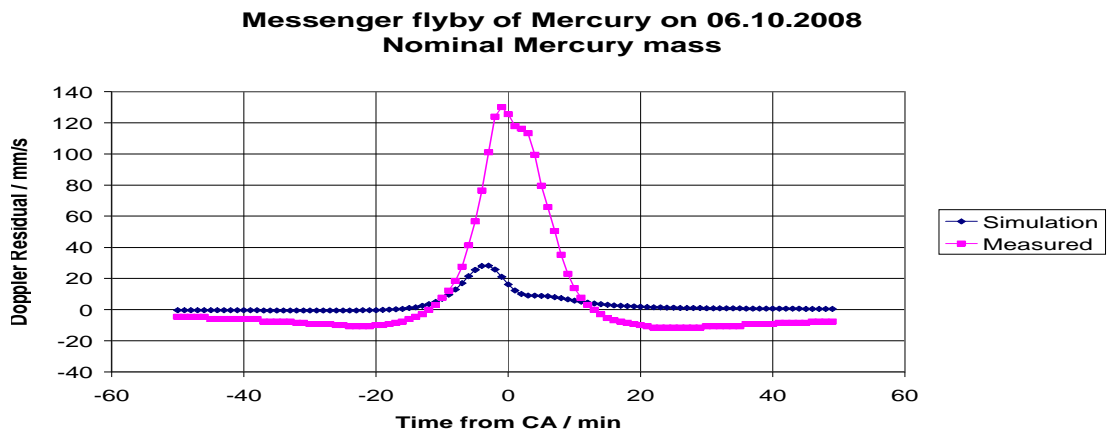

The residual perturbations of the simulated first flyby of Fig.1 are about 27mm/s and those of the second flyby of Fig.2 are about 30mm/s. This is in good congruence with the results of Smith et al..
Clearly, a final evaluation can not be done until the gravity field of Mercury is better estimated by the forthcoming orbits of Messenger.

But there is another way to make a closer comparison, even without knowledge of the details of the Mercury gravity field. As Smith et al. stated, the main effect in minimizing the Doppler residuals was obtained by increasing the mass of Mercury over that estimated by Mariner 10.
If the simulation of the flyby is of some evidence, then the calculation with an increased Mercury mass must have a significant gain in compliance with the measured data based on the uncorrected gravity data.
To do so, the mass was gained by a certain correction factor for the Newtonian track only, in order to simulate the conditions of the Doppler residual calculations by Smith et al. with the Mariner 10 data, namely the calculation with too low a mass for Mercury. Of course, this again leads to asymptotic residuals, again minimized by variation of the center coordinates of Mercury. Interestingly, now the needed variations were less than fifty meters.
The curves from Fig.3 and Fig.4 were derived with a correction factor of 1.00003.

**Fig.3** : Measured and simulated Doppler residuals of the second Messenger flyby of Mercury with increased mass. Measured values after [4] for a gravity field of Mercury from Mariner 10 data.

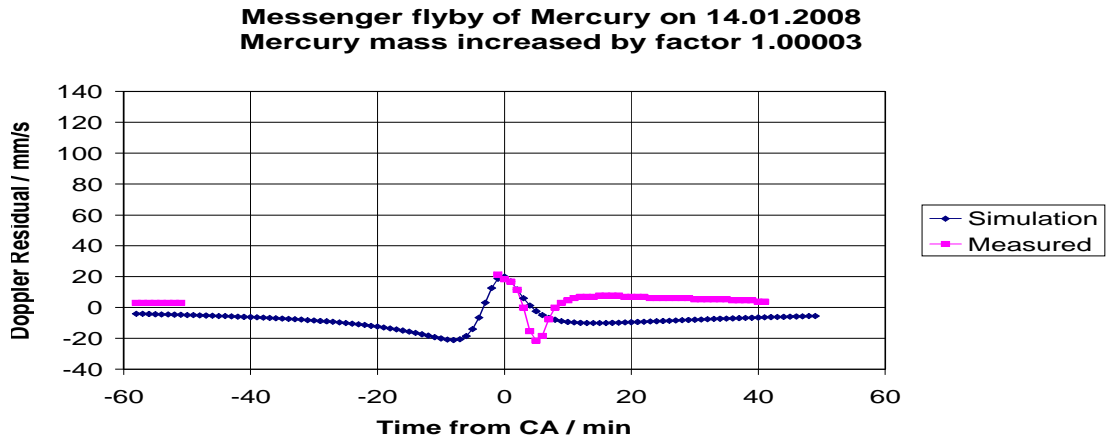

**Fig.4** : Measured and simulated Doppler residuals of the second Messenger flyby of Mercury with increased mass. Measured values after [4] for a gravity field of Mercury from Mariner 10 data.

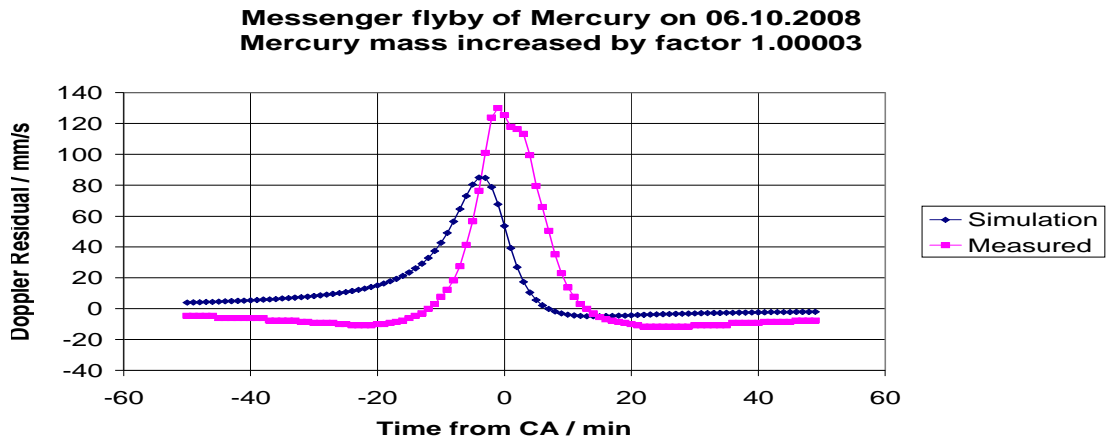

As one can see, the congruence with the measured data is significantly gained, as is to be expected with a correct simulation formula, if the statement is true, that increasing of the mass was the main effect in minimizing the residuals. Compared to the original mass data, the increased mass yields similar residual perturbation of about 30mm/s for the first flyby (in the window with comparable Doppler data), but much larger perturbations of about 90mm/s for the second flyby. Compared with the measured values of 43mm/s and 141mm/s this is an acceptable congruence, regarding the poor known Mercury gravity field.

Of course, there remain some differences of the simulated curve shape to the measured values. Smith et al. suspected in their summary, that, according to their calculations, Mercury will be seen to have mass concentrations like the Moon, able to explain the measured Doppler residuals, when Messenger enters Mercury orbit in 2011.
This would likely be true to a certain amount.
However, I suspect, according to this simulations, that this will fail to completely explain the doppler residuals. Possibly it will turn out, that there is indeed a flyby anomaly on both mercury flybys.

A further observation is remarkable. Mariner 10 has passed Mercury three times at perigee altitudes of about 703km, 48000km and 327km. The gravity field derived from these flybys is probably a weighted mean value of all three flybys, whereas the $2^{nd}$ degree coefficients mainly are derived from the third flyby.

The empirical formula (1a) with parameter set "low" leads for an altitude h of 327km to a mean additional value of about 9E-5 of the Mercury gravity field . For an altitude of about 200km, the perigee altitude of the Messenger spacecraft, the value is about 13E-5.

If the formula is correct to some extent, this means, that Messenger has seen a Mercury mass of about 1.00004 of the Mariner Mercury mass. This is very close to the value 1.00003 used for the above simulations with increased Mercury mass.

The parameter set "high" leads to different values but to roughly the same ratio of 1.00004.

In the introduction, the calculated flyby anomaly values for both Mercury flybys are given, derived from the formerly used parameter sets. Some remarks have to be made regarding these results.

The parameter sets are based on the data available in 2007. In the meantime, some of the measured anomaly values, especially those for the Galileo2 flyby and the Cassini flyby, obviously have been corrected. The formerly (2006) presented values were +0.11mm/s for Cassini and no measurable value for Galileo 2 [8]. For Galileo2, there was too much atmospheric drag due to the low perigee altitude. This drag could not be modelled with adequate accuracy, therefore masking any possible effect .

Recently, Anderson et al. [3] have published a value of $(-2 _{+/-1})$mm/s for Cassini and $(-4.6_{+/-1})$mm/s for Galileo 2. They gave no justification for the correction. These new values match well the results of the proposed Anderson formula, but now showing some discrepancies to the Busack simulation with a parameter set derived for the old values.

If there is no profound justification for that correction, the former values are to be considered as equally valuable as the new ones. The fact, that the Anderson formula matches well these values, is quite remarkable, but is somewhat relativized by the fact, that it failed to predict the following Rosetta flybys correctly.

So, in my opinion it seems to be possible, that the new data for Galileo2 and Cassini have to be further corrected, too. In this paper, therefore, no direct use of the new data has been made.

Instead, it was checked, whether it is possible with these data, to obtain a parameter set with a better match for the corrected values, while maintaining the values for the well observed flybys of Earth, especially the null results for the recent Rosetta flybys, in order to evaluate an error bar for the predicted Mercury flyby anomalys. In 2007, parameter sets with negative values for Galileo 2 and Cassini had been found, but were discarded because of poor match with the old data ([1, 16], equation (5)). Now, searching for new parameters for equation (1), not only a parameter set for the new data, but also an additional parameter set for the old values has been found. Both sets match roughly the respective anomaly values, while having apex coordinates, different from the former coordinates in opposite directions.

The result of this comparison was, that with these alternative parameter sets there is a big impact on the calculated Mercury flyby anomalies. Though all calculated values were negative with considerable magnitude, there were significant deviations from the nominal values given in the introduction, especially for the new Galileo2 and Cassini values. Therefore, no error limit has been specified. Based on the calculations with equation (1) , it only can be stated, that there must be a negative anomaly value with considerable magnitude of the order of several mm/s for both flybys , much more than predicted by the Anderson formula.

Interestingly, the calculated doppler residuals were without significant changes.

The results of the simulation presented here are based on a relatively complex program structure. Considerable effort has been made to find and to eliminate programming errors. The formal correctness of the results has been verified in a plurality of special tests. Nevertheless, it cannot be ruled out, that the achieved results are based on partly incorrect calculation. Considering the importance of the results in the case of correctness, it would be desirable to have the presented simulation reproduced by an independent party.

## 6  Conclusion

The empirical formula introduced by Busack [1] for simulating the observed flyby anomalies has been confirmed by the recent Rosetta flybys of Earth [2]. This success has motivated a test of the formula on the Doppler residuals of the first and second Messenger flybys of Mercury. For this test, the simulation program has been extended by adding the data of both Messenger flybys and the position vector of Mercury from Earth for the time of the flybys and by means for adjusting the center coordinates and the mass of Mercury. With these data, the Doppler velocity and range residuals were calculated without altering the previously derived parameter set, and compared with the data derived by Smith et al. [4]. The compliance was acceptable, regarding the lack of exact gravity data of Mercury, and the extrapolation of the Earth-derived parameter set to another planet. According to the formula and to the formerly derived parameter set, the expected flyby anomaly values were calculated. These predictions could be falsified or confirmed when the Mercury orbit of Messenger in 2011 will lead to better gravity data of the planet. If it would be confirmed then to an acceptable amount, the formula has described adequately 10 flyby anomaly values and additionally the curve shapes of the measured doppler residuals in the vicinity of closest approach with only 6 free parameters. This would be a strong indication of an underlying physical reality.